\documentclass[aps,pra,superscriptaddress,reprint,floatfix,longbibliography,showpacs,amsfonts,amsmath,amssymb]{revtex4-1}
\usepackage{graphicx}
\usepackage{hyperref}
\usepackage{upgreek}
\usepackage{epstopdf}
\usepackage{color}

\begin{document}
\title{Competition between spin-orbit coupling,
 magnetism, and dimerization in the
honeycomb iridates: $\alpha$-Li$_{2}$IrO$_{3}$ under pressure}

\author{V. Hermann}
\affiliation{Experimentalphysik II, Augsburg University, 86159 Augsburg, Germany}
\author{M.~Altmeyer}
\affiliation{Institut f\"ur Theoretische Physik, Goethe-Universit\"at
Frankfurt, 60438 Frankfurt am Main, Germany}
\author{J. Ebad-Allah}
\affiliation{Experimentalphysik II, Augsburg University, 86159 Augsburg, Germany}
\affiliation{Department of Physics, Tanta University, 31527 Tanta, Egypt}
\author{F.~Freund}
\author{A.~Jesche}
\author{A.~A.~Tsirlin}
\affiliation{Experimentalphysik VI, Center for Electronic Correlations and Magnetism,
Augsburg University, 86159 Augsburg, Germany}
\author{M.~Hanfland}
\affiliation{European Synchrotron Radiation Facility (ESRF), BP 220, 38043 Grenoble, France}
\author{P. Gegenwart}
\affiliation{Experimentalphysik VI, Center for Electronic Correlations and Magnetism,
Augsburg University, 86159 Augsburg, Germany}
\author{I.~I.~Mazin}
\affiliation{Code 6393, Naval Research Laboratory, Washington DC 20375, USA}
\author{D.~I.~Khomskii}
\affiliation{II. Physikalisches Institut, Universit\"at zu K\"oln, 50937 K\"oln, Germany}
\author{R.~Valent\'{\i}}
\email{valenti@th.physik.uni-frankfurt.de}
\affiliation{Institut f\"ur Theoretische Physik, Goethe-Universit\"at
Frankfurt, 60438 Frankfurt am Main, Germany}
\author{C. A. Kuntscher}
\email{christine.kuntscher@physik.uni-augsburg.de}
\affiliation{Experimentalphysik II, Augsburg University, 86159 Augsburg, Germany}

\begin{abstract}
Single-crystal x-ray diffraction studies with synchrotron radiation on the
honeycomb iridate $\alpha$-Li$_{2}$IrO$_{3}$ reveal a pressure-induced
structural phase transition with symmetry lowering from monoclinic to triclinic at a critical pressure of
$P_{c}$ = 3.8~GPa. According to the evolution of the lattice parameters with
pressure, the transition mainly affects the $ab$ plane and thereby the Ir
hexagon network, leading to the formation of Ir--Ir dimers. These observations
are independently predicted and corroborated by our \textit{ab initio} density
functional theory calculations where we find that the appearance of Ir--Ir
dimers at finite pressure is a consequence of a subtle interplay between
magnetism, correlation, spin-orbit coupling, and covalent bonding. Our results
further suggest that at $P_{c}$ the system undergoes a magnetic collapse.
Finally we provide a general picture of competing interactions for the
honeycomb lattices {$A_{2}$}$M$O$_{3}$ with $A$= Li, Na and $M$ = Ir, Ru.
\end{abstract}


\maketitle

In recent years, layered honeycomb 4$d$ and 5$d$ metal oxides, such as
Na$_{2}$IrO$_{3}$, $\alpha$-Li$_{2}$IrO$_{3}$, and $\alpha$-RuCl$_{3}$, have
been intensively scrutinized as Kitaev physics
candidates~\cite{Kitaev.2006,Jackeli.2009,Chaloupka.2010,Choi.2012,Plumb.2014,Winter.2017}
due to the presence of sizable nearest-neighbor bond-dependent spin-orbital
1/2 Ising interactions. However, instead of the expected $Z_{2}$ spin liquid
groundstate, as shown by Kitaev~\cite{Kitaev.2006}, these materials order
magnetically either in a zig-zag
structure~\cite{Choi.2012,Johnson.2015,Sears.2015,Banerjee.2016} (Na$_{2}%
$IrO$_{3}$, $\alpha$-RuCl$_{3}$) or an incommensurate spiral
structure~\cite{Williams.2016} ($\alpha$-Li$_{2}$IrO$_{3}$). This magnetic
long-range order has been suggested to originate from further non-Kitaev
interactions and a present debate is whether the magnetic excitations in these
materials nevertheless retain some of the non-trivial features of the Kitaev
model, such as
fractionalization~\cite{Banerjee.2016,Winter.2017a,Kasahara.2017}. It might be
expected that one route to enhance Kitaev interactions would be by applying
pressure or by doping. However, the physics of this structural family is much
richer and there are many more instabilities that interfere with the Kitaev
interactions, in particular under pressure. Indeed, Li$_{2}$RuO$_{3}$ is
nonmagnetic and strongly dimerized at ambient pressure
\cite{Miura.2007,Park.2016,Kimber.2014}, while SrRu$_{2}$O$_{6}$ is an
ultra-high-temperature antiferromagnet~\cite{Hiley.2015,Streltsov.2015},
despite having the same planar geometry, and shows no bond disproportionation.

Many factors control the competition between Kitaev physics, magnetism, and
dimerization~\cite{Streltsov.2017} in $A$$_{2}$$M$O$_{3}$ honeycomb networks,
such as the number of transition metal $M$ $d$-electrons, the strength of spin-orbit
coupling, the strength of correlation effects and Hund's rule coupling, or the
ionic radii of the buffer element $A$. In this context it is particularly
instructive to compare $\alpha$-Li$_{2}$IrO$_{3}$ with Li$_{2}$RuO$_{3},$
which contains the same buffer element (Li). $\alpha$-Li$_{2}$IrO$_{3}$ is less
correlated than Li$_{2}$RuO$_{3}$ (5$d$ versus 4$d$ electrons, resp.) so that one
would expect in the former a reduced tendency to magnetism in favor of
dimerization. On the other hand, $\alpha$-Li$_{2}$IrO$_{3}$ has only a single
hole in the $t_{2g}$ manifold, as opposed to two in Li$_{2}$RuO$_{3}$, and
stronger spin-orbit interaction. This should weaken dimerization and
strengthen Kitaev-type physics in $\alpha$-Li$_{2}$IrO$_{3}$. Further,
in  comparison  to Na$_{2}$IrO$_{3}$, the
Li ionic radius is smaller than the Na ionic radius, thus favoring
dimerization in Li$_{2}$$M$O$_{3}$. The result is a delicate balance of all these effects. At
ambient pressure, $\alpha$-Li$_{2}$IrO$_{3}$ shows a highly symmetric
honeycomb structure with a bond disproportionation of less than 3\% and
magnetically orders in a spiral structure, which features a noncollinear
incommensurate antiferromagnetic order inside Ir planes\cite{Williams.2016}.
It is not unreasonable to assume that $\alpha$-Li$_{2}$IrO$_{3}$ may be an
intermediate case between Na$_{2}$IrO$_{3}$ and Li$_{2}$RuO$_{3}$, and could
be switched between the two extremes (magnetic Kitaev and nonmagnetic
dimerized) by an external perturbation, such as physical pressure, an
intriguing possibility.

In this work, we show that this is indeed the case. We find that, in contrast
to Na$_{2}$IrO$_{3}$, in Li$_{2}$IrO$_{3}$ a structural phase transition from a monoclinic to a
dimerized triclinic structure occurs at a low pressure of $P_{c}$ = 3.8~GPa.
\textit{Ab initio} density functional theory (DFT) calculations also find this
transition, with the transition pressure depending on the assumed correlation strength.
The experimental $P_{c}$ is obtained for an effective Hubbard
repulsion interaction $U-J=1.5$~eV ($U$ and $J$ being the Hubbard and Hund's
rule coupling parameters, resp.), which is very reasonable for a 5$d$
metal. Analyzing the calculations we observe that indeed $\alpha$-Li$_{2}%
$IrO$_{3}$ is a borderline case, wherein the singlet dimerized solution and
the magnetic undimerized one are very close in energy. Pressure reduces the
tendency to magnetism, thus diminishing the energetic advantage of forming an
antiferromagnetic state, and brings Ir ions closer together, enhancing the
advantage of forming covalent bonds. Compared to Li$_{2}$RuO$_{3}$, the main
difference lies in the fact that in $\alpha$-Li$_{2}$IrO$_{3}$ only one
$d$-hole participates in the formation of covalent bonds.

At ambient pressure $\alpha$-Li$_{2}$IrO$_{3}$ crystallizes in a monoclinic
$C2/m$ space group \cite{Freund.2016,OMalley.2008}, with the unit cell shown
in the Supplemental Material~\cite{Suppl}. Ir forms hexagons with edge-sharing
IrO$_{6}$ octahedra and a single Li atom in its center (Ir$_{2}$LiO$_{3}$
layer). These layers are intercalated with Li$_{3}$O$_{3}$ layers. At ambient
pressure all Ir--Ir bonds have nearly the same length.

\begin{figure}[ptb]
\includegraphics[width=0.4\textwidth]{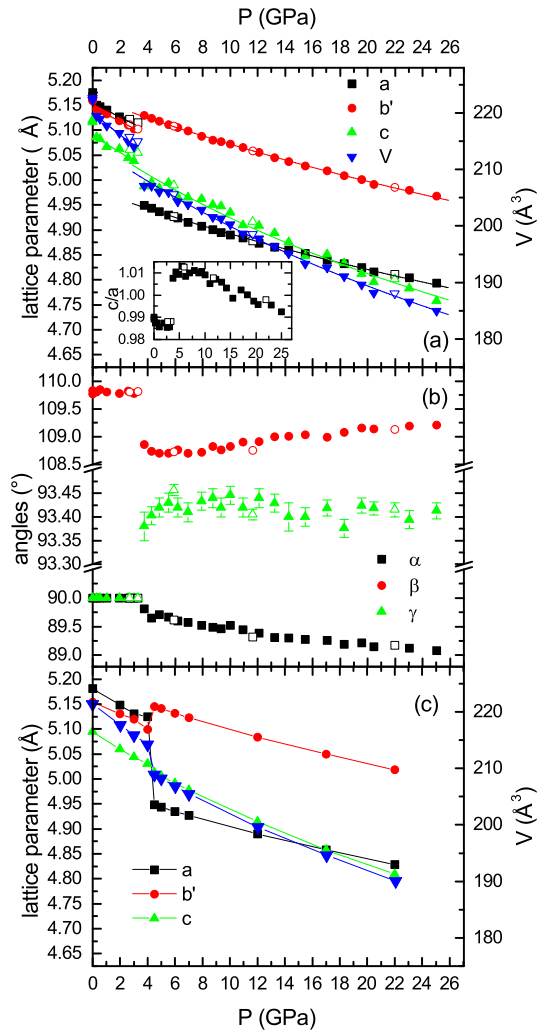}\caption{Pressure dependence
of the volume $V$ of the unit cell and the lattice parameters $a$, $b^{\prime
}=b/\sqrt{3}$, $c$ and $c/a$ value shown in (a) and the corresponding angles
$\alpha$, $\beta$ and $\gamma$ in (b). For a better comparison a non-primitive
unit cell was used for the high-pressure phase above 3.8~GPa, as explained in
the text. The solid lines in (a) are fits with a Murnaghan type equation of
state (see text). Open symbols are values observed during pressure release. If
not shown, the error bar is smaller than the symbol size. (c) Calculated
lattice parameters for pressures up to $22$~GPa (corrected by $3$~GPa to
closely resemble the experimental crystal volume at zero pressure).}%
\label{fig.1}%
\end{figure}

The pressure dependence of the lattice parameters obtained by refining the
single-crystal XRD data is depicted in Figs.~\ref{fig.1}(a)+(b). See the
Suppl. Material \cite{Suppl} for a description of sample preparation and
pressure-dependent XRD measurements. For pressures up to $\approx$3.8~GPa the
lattice parameters decrease monotonically with increasing pressure. We
included $b^{\prime}=b/\sqrt{3}$ for a better comparison between the three
lattice parameters. Both in-plane lattice parameters $a$ and $b^{\prime}$ are
affected in a very similar manner. The strongest effect is observed for the
lattice parameter $c$. This is illustrated by the $c/a$ value in the inset of
Fig.~\ref{fig.1}(a), which decreases with pressure up to 3.8~GPa. Within the
error bar the monoclinic angle $\beta$ is not affected by pressure.

For pressures above $P_{c}$=3.8~GPa the crystal structure can no longer be
refined within the $C2/m$ space group, but requires a triclinic group,
$P\overline{1},$ with half of the unit cell volume. The structural phase
transition is completely reversible [see open symbols in Figs.\ref{fig.1}%
(a)+(b)]. In order to compare the high-pressure evolution of the lattice
parameters to the low-pressure structures, we consider an supercell,
isometric to the ambient-pressure unit cell. Note that due to the triclinic
symmetry group, two different choices are possible. We select one of these
(resulting in the $X1$ dimerization discussed below) and show the refined lattice parameters
of this doubled triclinic unit cell in Figs.~\ref{fig.1}(a)+(b). All angles of
this cell exhibit an anomaly at $P_{c}$: $\gamma$ jumps from 90$^{\circ}$ to a
$\approx$93.4$^{\circ}$ and is independent of pressure above $P_{c}$ within
the error bars, while $\alpha$ monotonically decreases with pressure, and
$\beta$ exhibits an abrupt but small ($\approx$1$^{\circ}$) decrease at
$P_{c}$ and then monotonically increases with pressure. The structural phase
transition also induces abrupt changes in the lattice parameters. Whereas the
in-plane lattice parameter $a$ is most affected with an overall reduction of
about 3\%, the in-plane lattice constant $b$ slightly increases at $P_{c}$.
Interestingly, at the transition the $c/a$ ratio jumps up from $\approx$0.99
to 1.01, but above it monotonically decreases with pressure. Thus, also above
$P_{c}$ the lattice is more compressible along the $c$ direction.

We fitted the volume $V$ and the lattice parameters $r$ ($r$=$a$,$b$,$c$)
separately for the low- and high-pressure phases with the second-order
Murnaghan equation of state \cite{Murnaghan.1944}, to obtain the bulk moduli
$B_{0,V}$ and $B_{0,r}$:%

\begin{equation}
V(p)=V_{0}\cdot\lbrack(B_{0}^{\prime}/B_{0,V})\cdot p+1]^{-1/{B_{0}^{\prime}}}%
\end{equation}%
\begin{equation}
r(p)=r_{0}\cdot\lbrack(B_{0}^{\prime}/B_{0,r})\cdot p+1]^{-1/3{B_{0}^{\prime}%
}}%
\end{equation}
with $B_{0}^{\prime}$ fixed to 4. The results are summarized in Table~\ref{tab.B0}%
. For the low-pressure phase we find $B_{0,V}\approx$106(5)~GPa. The bulk
moduli $B_{0,a}$ and $B_{0,b}$ are almost the same, while  $B_{0,c}$ is
lower, confirming that $\alpha$-Li$_{2}$IrO$_{3}$ is more compressible along
the $c$ direction. Above $P_{c}$ the bulk modulus is increased to
$B_{0,V}\approx$125(3)~GPa, while $B_{0,a}$ and $B_{0,b}$ sharply increase to
$\approx$170~GPa, and $B_{0,c}$ is slightly decreased compared to the
low-pressure phase to about 86(5)~GPa. Thus, $\alpha$-Li$_{2}$IrO$_{3}$
hardens at $P_{c}$, and the hardening takes place within the Ir planes.

\begingroup
\squeezetable
\begin{table}[b]
\caption{Bulk moduli $B_{0,V}$ and $B_{0,r}$ with $r$=$a$,$b$,$c$ of $\alpha
$-Li$_{2}$IrO$_{3}$ in the low-pressure (low-p) and high-pressure (high-p)
phase, as obtained from fitting the volume $V$ with a Murnaghan-type equation
of state, with $B_{0}^{\prime}$ set to 4. Corresponding values for Na$_{2}%
$IrO$_{3}$ were obtained from corresponding fits of the data given in
Ref.~\cite{Hermann.2017}.}%
\label{tab.B0}
\begin{ruledtabular}
\begin{tabular}{rccccc}
& $B_{0,V}$ & $V_0$ [\AA$^3$] & $B_{0,a}$ [GPa] & $B_{0,b}$ [GPa] & $B_{0,c}$ [GPa]\\
low-p phase & 106(5) & 220.1(2) & 114(2) & 113(2) & 92(11) \\
high-p phase & 125(3) & 214.3(9) & 167(2) & 166(2) & 86(5) \\
Na$_2$IrO$_3$ & 100.6(1) & 269.55(3) & 152(2) & 146 (2) & 58(1)
\end{tabular}
\end{ruledtabular}
\end{table}\endgroup

For a comparison with isostructural Na$_{2}$IrO$_{3}$ we refitted the data of
Ref.~\cite{Hermann.2017} with $B_{0}^{\prime}$ fixed to 4 and list the
so-obtained results in Table~\ref{tab.B0}. Compared to $\alpha$-Li$_{2}%
$IrO$_{3}$ in the low-pressure phase, the bulk modulus $B_{0,V}$ of Na$_{2}%
$IrO$_{3}$ is only slightly lower. Interestingly, for Na$_{2}$IrO$_{3}$ the
values $B_{0,a}$ and $B_{0,b}$ are much higher, whereas $B_{0,c}$ is much
lower. Hence, in its low-pressure phase $\alpha$-Li$_{2}$IrO$_{3}$ is more
compressible in the $ab$ plane and less compressible along the $c$ direction
as compared to Na$_{2}$IrO$_{3}$. The former effect could be attributed to the
smaller Li atoms occupying the center of the hexagons (instead of Na), and the
latter to the stronger Li--O bonds with strong out-of-$ab$-plane
character~\cite{Hermann.2017}. The higher in-plane compressibility is
instrumental in triggering Ir dimerization. Above $P_{c}$ the bulk moduli
$B_{0,a}$ and $B_{0,a}$ of $\alpha$-Li$_{2}$IrO$_{3}$ are sharply enhanced and
become similar to those for Na$_{2}$IrO$_{3}$, \textit{i.e.}, the
compressibility of the $ab$ plane becomes similar for both compounds.

\begin{figure}[ptb]
\includegraphics[width=.4\textwidth]{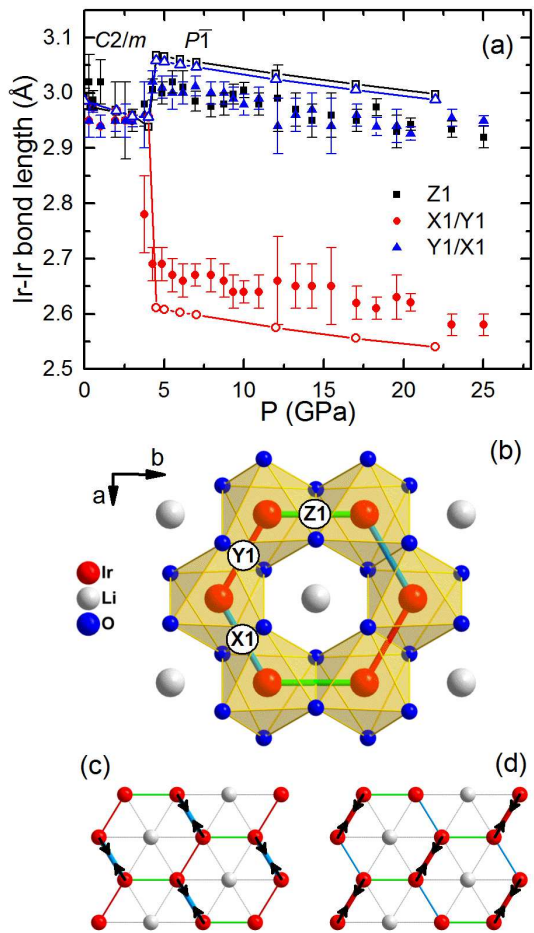}\caption{(a) Pressure
dependence of the Ir--Ir distances for the Ir hexagons in the $ab$ plane, with
the nomenclature (Ir bonds $X1$, $Y1$, $Z1$) given in (b) for the
ambient-pressure monoclinic crystal structure. Ir--Ir bond lengths as
predicted from our \textit{ab initio} relaxations are shown as open symbols
connected by a line. The two equivalent ordering patterns of the Ir--Ir dimers
above $P_{c}$ are illustrated in (c) and (d).}%
\label{fig.2}%
\end{figure}

For a more detailed investigation, we refined the Ir--Ir bond lengths
[see Fig.~\ref{fig.2} (a)], since they are relevant for the magnetic and
electronic properties of the
material~\cite{Winter.2017,Nishimoto.2016,Winter.2016}. In a perfectly
hexagonal lattice one can distinguish three Ir--Ir bonds related by a 120$%
{{}^\circ}%
$ rotation [Fig.~\ref{fig.2} (b)]. Following the nomenclature of
Ref.~\cite{Winter.2016}, we shall call them $Z1$, $X1$, and $Y1$. In the
monoclinic phase below $P_{c}$, the $X1$ and $Y1$ bonds are equivalent by
symmetry, while $Z1$ is distinct. The high-pressure phase lacks the $C_{2}$
symmetry, and therefore the two bonds $X1$ and $Y1$ become inequivalent. The
Ir--Ir bond lengths as a function of pressure, as obtained from the refinement
of the XRD data, are plotted in Fig.~\ref{fig.2}(a). At $P_{c}$ one of the $X1/Y1$ bonds is slightly increased
from $\approx$2.95~{\AA } to $\approx$3.00~{\AA }, while the other one is
\textit{strongly decreased} to 2.69~{\AA }. Note that this distance is smaller
than the Ir--Ir distance of 2.714 {\AA } in Ir metal. As opposed to Li$_{2}%
$RuO$_{3},$ where (i) the dimerized bonds alternate between $X1$ and $Y1,$
 (ii) a
$C_{2}$ axis is preserved and (iii) the $P2_{1}/m$ monoclinic symmetry is realized,
in $\alpha$-Li$_{2}$IrO$_{3}$ either $X1$ or $Y1$ bonds dimerize, thus
breaking the $C_{2}$ symmetry [Figs.~\ref{fig.2} (c,d)]. The $Z1$ bond's length
increases at $P_{c}$ and becomes nearly degenerate with that of the longer
$X1/Y1$ bond. The energy difference between these various types of
dimerization is related to different mutual arrangements of the
dimers: armchair (herringbone) or ladder (parallel). As dicussed in Refs.
~\cite{Kimber.2014,Jackeli.2008} the choice is being made by long-range,
likely elastic, interactions. It is worth noting that very recently Veiga
\textit{et al.}~\cite{Veiga.2017} have observed a structural phase transition
at $P_{c}\approx4$ GPa in a similar 5$d$-system $\beta$-Li$_{2}$IrO$_{3}$, which
is consistent with dimerization above $P_{c}$, although no actual evidence of
dimerization was obtained.

In order to gain more insight into the physics of dimerization and
interactions that control it, we have performed first-principles DFT
calculations, as described in the Suppl. Material~\cite{Suppl}. First, we
find, in agreement with experiment and qualitative considerations, that
$\alpha$-Li$_{2}$IrO$_{3}$ is more prone to dimerization than Na$_{2}$%
IrO$_{3},$ and, in fact, dimerizes in DFT within the generalized gradient
approximation (GGA) already at zero pressure. This can be traced to the
underestimation of correlation effects, and, therefore, underestimation of the
tendency to magnetism. Indeed, Ref.~\cite{Petukhov.2003} showed that at the
mean-field level, usually called the LDA+U approximation (or GGA+U), correlations
increase the Stoner coupling $I$, which characterizes the tendency to form
magnetic moments in DFT, as $I\rightarrow I+(U-J)/5,$ where $U$ and $J$ are
the Hubbard and Hund's rule coupling parameters. For 5$d$ metals $I$ is of the
order of 0.3~eV, while $U_{\text{eff}}=U-J,$ as we discuss in more detail in
the Suppl. Material~\cite{Suppl}, should be taken to be close to $1.5$~eV.
Thus, including correlations enhances magnetic interactions by about a factor
of two.

Furthermore, in accordance with our previous study~\cite{Manni.2014} for the
isoelectronic doping series Na$_{(1-x)}$Li$_{x}$IrO$_{3}$, we find that at
ambient pressure, using the experimental crystal volume, it is not enough
to include an effective Hubbard correlation of about $U_{\text{eff}}=U-J=1.5$~eV
 to stabilize magnetism and the observed undimerized structure.
The spin-orbit coupling also appears essential. Here we choose the
 zigzag antiferromagnetic order to mimic the
incommensurate spiral order~\cite{note1}
observed in experiment~\cite{Williams.2016}. This indicates that the observed
structural transition at finite pressures is driven by two competing energy
scales: (i) the energy gained by the formation of magnetic moments and (ii)
the energy gained by placing two electrons of neighboring Ir atoms in the
bonding orbital resulting from the Ir--Ir dimerization. In both processes,
correlation effects are significantly involved. At low pressures the overlap
of adjacent Ir atoms is slightly too small to favor the dimerization, while a
small decrease of the Ir--Ir distances with pressure changes the situation and
results in a breakdown of the magnetic order and structural dimerization. Note
that in the above-mentioned experiment on $\beta-$ Li$_{2}$IrO$_{3}%
$~\cite{Veiga.2017} static magnetism was disappearing prior to the putative
dimerization transition, at $P_{m}\sim2$~GPa, and one can speculate that at a
higher pressure, corresponding to dimerization, the local moments collapse as
well, in agreement with the present calculations. Alternatively, between $P_{m}$ and
$P_{c}$ the system may be in a valence bond liquid state, similar to that in
Li$_{2}$RuO$_{3}$ at high temperature\cite{Kimber.2014}.

In Fig.~\ref{fig.1} (c) we show the lattice parameters calculated within
relativistic GGA+U as a function of the simulated hydrostatic pressures. As
typical for GGA, we observe slight underbinding, so that the experimental
volume at ambient pressure corresponds to the calculated pressure of
$\approx3$~GPa. In the following, we subtract this systematic error from the
calculated pressure. With this in mind, the calculated transition pressure as
well as the resulting bond disproportionations are in good agreement with the
experiment. At the phase transition, the $X1/Y1$ dimerized structure is the
most stable with a relative Ir--Ir dimerization of long/short \thinspace
bond$=3.07/2.61\approx1.18$ in very good agreement with the experiment, as
shown in Fig.~\ref{fig.2}(a). In addition, we find an abrupt collapse of
magnetism at $P_{c}$ (see Suppl. Material~\cite{Suppl}).

At this point, it is insightful to investigate the dimerization pressure for
the closely related Na$_{2}$IrO$_{3}$ compound. In fact, at $29$~GPa the
crystal structure of Na$_{2}$IrO$_{3}$ has been shown to resemble that of
$\alpha$-Li$_{2}$IrO$_{3}$ with respect to the lattice metrics \cite{Hermann.2017}. Our simulations for Na$_{2}%
$IrO$_{3}$ (see Suppl. Material~\cite{Suppl}) predict lattice parameters that
are in excellent agreement with the observations in the experimentally
studied pressure range up to $30$~GPa. At about $45$~GPa, we find a
structural transition perfectly analogous to the one we observed in $\alpha
$-Li$_{2}$IrO$_{3}$. It is accompanied by a magnetic collapse and the Ir--Ir
dimerization triggers a strong decrease of the in-plane $a$ lattice parameter,
while the $b$-lattice parameter is slightly enhanced. The $c$ parameter in
Na$_{2}$IrO$_{3}$ is significantly larger than in $\alpha$-Li$_{2}$IrO$_{3}$,
due to the larger intercalated ion in Na$_{2}$IrO$_{3}$. Notably, the Ir--Ir
distance, at which the transition occurs, about $2.95$~$\text{\AA ,}$ is
almost exactly the same as for $\alpha$-Li$_{2}$IrO$_{3}$ and also the unit
cell volume is remarkably similar ($V\approx210\text{\AA }^{3}$). Indeed, due
to the larger Na ion size compared to Li, the IrO$_{2}$ layer is less
compressible in Na$_{2}$IrO$_{3}$, as discussed above, and thus formation of
sufficiently short dimers (in order to take full advantage of the covalent
energy) is hindered. This is also validated by our bulk modulus study (Table
\ref{tab.B0}). Otherwise, as expected due to their similar electronic
properties, these two compounds behave in the same way. The pressure of
$\approx45$~GPa, at which we expect the dimerization transition in
Na$_{2}$IrO$_{3}$ to occur, should be accessible in diamond anvil cells but
is beyond the scope of the present study. Conducting such an experiment in the
future will be invaluable to confirm the general physical picture that we
deduce in this paper.

In summary, $\alpha$-Li$_{2}$IrO$_{3}$ undergoes a pressure-induced structural
phase transition at $P_{c}$=3.8~GPa with symmetry lowering to $P\overline{1}$.
This transition mainly affects the $ab$ plane with the Ir hexagons. As
corroborated by our density functional theory calculations, our refinements of
the Ir positions show that the structural phase transition is accompanied by a
dimerization of the previously equally long $X1/Y1$ Ir--Ir bonds. Analysis of
the total energies of the high-symmetry and the dimerized $P\overline{1}$
lattices at the experimental crystal volume shows that several factors affect
the propensity to dimerization, namely the size of the central ion
(\textit{i.e., }Li \textit{vs.} Na), as well as the strength of the spin-orbit
interaction, electronic correlations, and Hund's rule coupling (all of them discourage
dimerization, \textit{cf. }4d \textit{vs. }5$d$ metals), and the number of
$d$-electrons in the metal species (\textit{e.g., }Ru \textit{vs. }Ir).
Li$_{2}$RuO$_{3},$ having (a) a small central ion, (b) four $d$-electrons, and
(c) weaker spin-orbit coupling has the strongest tendency towards dimerization
among the known 213 systems, despite being more magnetic than Ir. $\alpha
$-Li$_{2}$IrO$_{3}$ has one more $d$-electron and stronger spin-orbit
interaction, but weaker Hubbard and Hund's rule couplings, and thus dimerizes
only under pressure, albeit at a relatively small $P_{c}$. Na$_{2}$IrO$_{3}$
only differs in terms of the central ion size, and thus compressibility, and
is predicted to dimerize as well, at an accessible, but much higher pressure.

\vspace*{1em}

We thank the ESRF, Grenoble, France, for the provision of beamtime. This work
was financially supported by the Federal Ministry of Education and Research
(BMBF), Germany, through Grant No. 05K13WA1 (Verbundprojekt 05K2013,
Teilprojekt 1, PT-DESY). DKh is grateful to D. Haskel for useful discussions.
MA, DKh, RV and PG acknowledge financial support by the Deutsche
Forschungsgemeinschaft (DFG), Germany, through TRR 80, SPP 1666, TRR 49 and
SFB 1238. AJ acknowledges support from the DFG through Grant No. JE 748/1. AAT
acknowledges financial support from the Federal Ministry for Education and
Research via the Sofja-Kovalevskaya Award of Alexander von Humboldt
Foundation, Germany. IIM was supported by A. von Humboldt foundation
and by ONR through the NRL basic research program.


\begin{thebibliography}{30}
\bibitem [{not()}]{note1}%
  \BibitemOpen
  \bibfield  {title} {\enquote {\bibinfo {title} {We note that without a
  further increase of the unit cell, only three antiferromagnetic patterns are
  possible: Neel ($\mathbf{q=}$0), stripes, and zigzag (both
  $\mathbf{q=}\{1,0\}$ in plane), where the latter two are closer to the
  experimental wave vector. the choice between stripes and zigzag is rather
  arbitrary, and does not affect the final conclusions.}}\ }\href@noop {} {\
  }\BibitemShut {NoStop}%
\end{thebibliography}
\end{document}